\documentclass[runningheads]{llncs}
\usepackage{graphicx}
\usepackage{subcaption}
\usepackage{listings}
\usepackage{csquotes}
\usepackage{appendix}
\usepackage{longtable}
\usepackage{algorithm,algorithmic}
\usepackage{mathtools}
\usepackage{amsmath}
\usepackage{color}
\usepackage{tcolorbox}
\usepackage{url}

\newtheorem{mydef}{Definition}

\newtheorem{myexc}{Exception}

\newtheorem{mymod}{Mode}

\makeatletter
\RequirePackage[bookmarks,unicode,colorlinks=true]{hyperref}%
   \def\@citecolor{blue}%
   \def\@urlcolor{blue}%
   \def\@linkcolor{blue}%

\def\orcidID#1{\smash{\href{http://orcid.org/#1}{\protect\raisebox{-1.25pt}{\protect\includegraphics{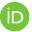}}}}}
\makeatother

\begin{document}
\title{An Efficient Floating-Point Bit-Blasting API for Verifying C Programs} 
%
%
\author{Mikhail R. Gadelha\inst{1}\orcidID{0000-0001-6540-6587} \and
Lucas C. Cordeiro\inst{2}\orcidID{0000-0002-6235-4272} \and
Denis A. Nicole\inst{3}
}
\authorrunning{M. Gadelha et al.}
%
\institute{
  SIDIA Instituto de Ci\^encia e Tecnologia, Manaus, Brazil \\ \email{mikhail.gadelha@sidia.com} \and
  University of Manchester, Manchester, UK\\ \email{lucas.cordeiro@manchester.ac.uk} \and
  University of Southampton, UK,\\ \email{dan@ecs.soton.ac.uk}
}
\maketitle              
\begin{abstract}
We describe a new SMT bit-blasting API for floating-points and evaluate it using different out-of-the-shelf SMT solvers during the verification of several C programs. The new floating-point API is part of the SMT backend in ESBMC, a state-of-the-art bounded model checker for C and C++. For the evaluation, we compared our floating-point API against the native floating-point APIs in Z3 and MathSAT. We show that Boolector, when using floating-point API, outperforms the solvers with native support for floating-points, correctly verifying more programs in less time. Experimental results also show that our floating-point API implemented in ESBMC is on par with other state-of-the-art software verifiers.  Furthermore, when verifying programs with floating-point arithmetic, our new floating-point API produced no wrong answers.
\end{abstract}
%
%
\section{Introduction}

Software verification tools operate by converting their input (e.g., a program source code) into a format understandable by an automated theorem prover, encoding high-level program properties (e.g., arithmetic overflow) and algorithms (e.g., bounded model checking) into low-level equations (e.g., SMT). The encoding process of a program usually involves several intermediate steps, designed to generate a formula that can be efficiently solved by the theorem provers. In this domain, the analysis of programs with floating-point arithmetic has received much attention, primarily when safety depends on the correctness of these programs. In essence, the Ariane 5 rocket exploded mid-air in 1996 due to an exception thrown by an invalid floating-point conversion~\cite{Ariane5}. It is a complex problem because the semantics may change beyond code level, including the optimization performed by compilers~\cite{Monniaux:2008:PVF:1353445.1353446}.

There exist various static analysis tools that are able to check for floating-point computations~\cite{abs-cs-0701193,Botella:2006:SEF:1133626.1133628,Clarke04,Collavizza:2014:GTC:2593735.2593737,Fu:2017:AHC:3062341.3062383,Quan:2016:HSE:2950290.2983966,Tillmann:2008:PWB:1792786.1792798}. For example, Astr\'ee  is a static analysis tool that considers all possible rounding errors when verifying C programs with floating-point numbers~\cite{abs-cs-0701193}. It has been applied to verify embedded software in the flight control software of the Airbus. CBMC~\cite{Clarke04} is also another notable example of a software model checking tool, which implements a bit-precise decision procedure for the theory of floating-point arithmetic~\cite{Brain2014}. It has been applied to verify industrial applications from the automotive industry, which rely on floating-point reasoning~\cite{DBLP:journals/corr/SchrammelKBMTB14}. CBMC is also the main verification engine employed by other software verifiers that efficiently verify C programs with floating-point numbers such as PeSCo~\cite{RichterW19} and VeriAbs~\cite{ChimdyalwarDCSK17}. It is a challenging verification task to prove the correctness of C programs with floating-points mainly because of $32$/$64$ bits floating-point computations. Given the current knowledge in software verification, there exists no other study that shows a thorough comparative evaluation of software verifiers and SMT solvers concerning the verification of C programs that contain floating-points.

Here we present the new floating-point technologies developed in one bounded model checker, ESBMC~\cite{esbmc2019}, and evaluate it using a large set of floating-point benchmarks~\cite{svcomp2020}. In particular, we evaluate a new floating-point API on top of our SMT backend that extends the floating-point feature to all solvers supported by ESBMC (including Boolector~\cite{Brummayer:2009:BES:1532891.1532912} and Yices~\cite{DBLP:conf/cav/2014} that currently do not support the SMT \texttt{FP} logic~\cite{7203811}). For evaluation, we used the benchmarks of the 2020 International Competition on Software Verification (SV-COMP)~\cite{svcomp2020}, from the floating-point sub-category. The five different solvers supported by ESBMC were evaluated (Z3~\cite{Z08}, Yices~\cite{DBLP:conf/cav/2014}, Boolector~\cite{Brummayer:2009:BES:1532891.1532912}, MathSAT~\cite{Cimatti:2013:MSS:2450387.2450400}, and CVC4~\cite{Barrett:2011:CVC:2032305.2032319}) and ESBMC is able to evaluate more benchmarks within the usual time and memory limits ($15$ minutes and $15$GB, respectively) when using Boolector. In particular, results show that Boolector can solve more floating-point problems using the new floating-point API than MathSAT or Z3, which have native floating-point APIs. Our experimental results also show that our floating-point API implemented in ESBMC is competitive to other state-of-the-art software verifiers, including CBMC~\cite{Clarke04}, PeSCo~\cite{RichterW19}, and VeriAbs~\cite{ChimdyalwarDCSK17}.

\section{Floating-point Arithmetic}
\label{back:fp}

The manipulation of real values in programs is a necessity in many fields, e.g., scientific programming~\cite{Monniaux:2008:PVF:1353445.1353446}. The set of real numbers, however, is infinite, and some numbers cannot be represented with finite precision, e.g., irrational numbers. Over the years, computer manufacturers have experimented with different machine representations for real numbers~\cite{1676076}. The two fundamental ways to encode a real number are the fixed-point representation, usually found in embedded microprocessors and microcontrollers~\cite{texas-dsp}, and the floating-point representation, in particular, the IEEE floating-point standard (IEEE 754-2008~\cite{4610935}), which has been formally accepted by many processors~\cite{Goldberg91whatevery}.

Each encoding can represent a range of real numbers depending on the word-length and how the bits are distributed. A fixed-point representation of a number consists of an integer component, a fractional component, and a bit for the sign. In contrast, the floating-point representation consists of an exponent component, a significand component, and a bit for the sign. Floating-point has a higher dynamic range than fixed-point (e.g., a \texttt{float} in C has $24$ bits of precision, but can have values up to $2^{128}$), while fixed-point can have higher precision than floating-point~\cite{Nikolic2007}. Furthermore, the IEEE floating-point standard defines values that have no equivalent in a fixed-point or real encoding, e.g., positive and negative infinities. In general, IEEE floating-points are of the following kinds: \textit{zeroes}, \textit{NaNs}, \textit{infinities}, \textit{normal}, \textit{denormal (or subnormal)}~\cite{4610935}.

\begin{mydef}
\label{Infinities}
\textbf{(Infinities)} Both \texttt{+inf} and \texttt{-inf} are defined in the
standard. These floating-points represent overflows or the result of non-zero
floating-point divisions by zero (Annex F of the C language
specification~\cite{C11}).
\end{mydef}

\begin{mydef}
\label{Zeroes}
\textbf{(Zeroes)} Both \texttt{+0} and \texttt{-0} are defined in the standard.
Most of the operations will behave identically when presented with \texttt{+0}
or \texttt{-0} except when extracting the sign bit or dividing by zero (usual
rules about signedness apply and will result in either \texttt{+inf} or
\texttt{-inf}). Equalities will even be evaluated to true when comparing
positive against negative zeroes.
\end{mydef}

\begin{mydef}
\label{NANs}
\textbf{(NaNs)} The \textbf{N}ot \textbf{a} \textbf{N}umber special values
represent undefined or unrepresentable values, e.g., $\sqrt{-1}$ or
\texttt{0.f/0.f}. As a safety measure, most of the operations will return NaN if
at least one operator is NaN, as a way to indicate that the computation is
invalid. NaNs are not comparable: except for the not equal
operator (\texttt{!=}), all other comparisons will evaluate to false (even comparing
a NaN against itself). Furthermore, casting NaNs to integers is undefined behavior.
\end{mydef}

\begin{mydef}
\textbf{(Normal)} A non-zero floating-point that can be
represented within the range supported by the encoding.
\end{mydef}

\begin{mydef}
\textbf{(Denormal (or subnormal))} A non-zero floating-point
representing values very close to zero, filling the gap between what can be usually
represented by the encoding and zero.
\end{mydef}

The IEEE standard also defines five kinds of exceptions, to be raised under
specific conditions, which are: \textit{invalid operation}, \textit{overflow},
\textit{division by zero}, \textit{underflow}, and \textit{inexact}~\cite{4610935}.

\begin{myexc}
\textbf{(Invalid Operation)} This exception is raised when the operation produces
a NaN as a result.
\end{myexc}

\begin{myexc}
\label{overflow}
\textbf{(Overflow)} This exception is raised when the result of an operation is
too large to be represented by the encoding. By default, these operations return
$\pm$\texttt{inf}.
\end{myexc}

\begin{myexc}
\textbf{(Division By Zero)} It is raised by \texttt{x/$\pm$0}, for
\texttt{x$\neq$0}. By default, these operations return $\pm$\texttt{inf}.
\end{myexc}

\begin{myexc}
\label{underflow}
\textbf{(Underflow)} Raised when the result is too small to be represented by the
encoding. The result is a denormal floating-point.
\end{myexc}

\begin{myexc}
\textbf{(Inexact)} This exception is raised when the encoding cannot represent
the result of an operation unless it is rounded. By default, these operations
will round the result.
\end{myexc}

The standard defines five rounding modes. Given a real number $x$, a rounded
floating-point $r(x)$ will be rounded using: \textit{Round Toward Positive (RTP)},
\textit{Round Toward Negative (RTN)}, \textit{Round Toward Zero (RTZ)},
\textit{Round to Nearest ties to Even (RNE)}, and
\textit{Round to Nearest ties Away from zero (RNA)}:

\begin{mymod}
\textbf{(RTP)} $r(x)$ is the least floating-point value $\geq x$.
\end{mymod}

\begin{mymod}
\textbf{(RTN)} $r(x)$ is the greatest floating-point value $\leq x$.
\end{mymod}

\begin{mymod}
\textbf{(RTZ)} $r(x)$ is the floating-point with the same
sign of $x$, such that $|r(x)|$ is the greatest floating-point value $\leq |x|$.
\end{mymod}

\begin{mymod}
\textbf{(RNE)} $r(x)$ is the floating-point value closest to
$x$; if two floating-point values are equidistant to $x$, $r(x)$ is the one
which the least significant bit is zero.
\end{mymod}

\begin{mymod}
\textbf{(RNA)} $r(x)$ is the floating-point value
closest to $x$; if two floating-point values are equidistant to $x$, $r(x)$ is
the one further away from zero.
\end{mymod}

The standard also defines some arithmetic operations (add, subtract, multiply, divide, square root, fused multiply-add, remainder), conversions (between formats, to and from strings), and comparisons and total ordering. In particular, the standard defines how floating-point operations are to be encoded using bit-vectors. Table~\ref{table:back-fp} shows four primitive types usually available in the x86 family of processors that follow the standard; each type is divided into three parts: one bit for the sign, an exponent, and a significant part which depends on the bit length of the type. The significands also include a hidden bit: a 1 bit that is assumed to be the leading part of the significand, unless the floating-point is denormal.
\begin{table}[!h]
\centering
\begin{tabular}{|l|c|c|c|}
 \hline
 \textbf{Name} & \textbf{Common Name} & \textbf{Size}           \\
    &       & \textbf{(exponent + significand)} \\
 \hline
 fp16 & Half precision    & 16 (5 + 10)  \\ \hline
 fp32 & Single precision   & 32 (8 + 23)  \\ \hline
 fp64 & Double precision   & 64 (11 + 53)  \\ \hline
 fp128 & Quadruple precision & 128 (15 + 113) \\ \hline
 \end{tabular}
 \caption{IEEE floating-point types.}
 \label{table:back-fp}
\end{table}

In Annex F of the C language specification~\cite{C11}, fp32 and fp64 are defined as \texttt{float} and \texttt{double}. The standard does not define any types for fp16, and compilers usually implement two formats: \texttt{\_\_fp16} as defined in the ARM C language extension (ACLE)~\cite{ACLE} and \texttt{\_Float16} as defined by the ISO/IEC 18661-3:2015 standard~\cite{ISO18661-3:2015}. While \texttt{\_\_fp16} is only a storage and interchange format (meaning that it is promoted when used in arithmetic operations), \texttt{\_Float16} is an actual type, and arithmetic operations are performed using half-precision. The standard only weakly specifies how an fp128 (\texttt{long double} in C) should be implemented, and compilers usually implement it using an 80-bit long double extended precision format~\cite{Goldberg91whatevery}.

Floating-points are represented as $(-1)^{sign} \times significand \times 2^{exponent}$. Here, $1 \leq significand \leq 2$ and $2^{exponent}$ is the scaling factor~\cite{0028245}. Regular floating-points are encoded assuming that the leading hidden bit is 1 and the exponent is in the range $[-exponent_{max}+1, exponent_{max}]$, e.g., the number 0.125 is represented as $\langle0011000000000000\rangle$ in the floating-point format. Denormals are encoded assuming that the leading hidden bit is zero and the exponent is $-exponent_{max}$. Zeros are represented as an all-zero bit-vector (except for the sign bit if the zero is negative). Finally, a bit-vector with the exponent equal to $exponent_{max}$ and significand all zero is an infinity. In contrast, a bit-vector with an exponent equal to $exponent_{max}$ and significand not zero is a NaN.

\section{A Floating-Point Bit-Blasting API for Verifying C Programs}

When ESBMC was created, all floating-point types and operations were encoded using fixed-points~\cite{AbreuGCFS16,BessaICF16,BessaIPCF17,ChavesBCKF17,IsmailBCFF15}. A fixed-point number is represented in ESBMC as a pair $(m,n)$ where $m$ is the total number of bits and $n \leq m$ is the number of fractional bits, e.g., the number 0.125 is represented as $\langle0000.0010\rangle$ (assuming it is 8 bits long) in the fixed-point format. The fixed-point arithmetic is performed similarly to the bit-vector arithmetic, except that the operations are applied separately to the integral and fractional parts of the fixed-points and concatenated at the end (overflow in the fractional parts are treated accordingly). Different from floating-points, all bit-vectors represent one number in the real domain.

The lack of proper floating-point encoding, however, meant that ESBMC was unable to accurately verify an entire class of programs, such as the famous floating-point ``issue''~\cite{famous-fp-issue} illustrated in Figure~\ref{fig:simple-fp}.
\begin{figure}[!ht]
 \centering
 \begin{subfigure}{.59\textwidth}
 \begin{lstlisting}[escapechar=^]
int main()
{
 double x = 0.1;
 double y = 0.2;
 double w = 0.3;
 double z = x + y;
 assert(w == z);^\label{nan-assert}^
 return 0;
}
\end{lstlisting}
 \end{subfigure}
\caption{The assertions in line~\ref{nan-assert} does
not hold when using floating-point arithmetic.}
\label{fig:simple-fp}
\end{figure}

The assertion in line~\ref{nan-assert} holds if the program is encoded using fixed-point arithmetic, but fails if floating-point arithmetic is used. The assertion violation arises from the fact that floating-points in the IEEE standard are represented as whole numbers $\times$ a power of two, so the only numbers that use a prime factor of the base two that can be correctly expressed as fractions. Since in binary (or base 2) the only prime factor is 2, only $\frac{1}{2}$, $\frac{1}{4}$, $\frac{1}{8}$, $\ldots$ would be correctly expressed as decimals, so the constants $0.1$, $0.2$, $0.3$ (or $\frac{1}{10}$, $\frac{1}{5}$, $\frac{1}{3}$) in the program are only approximations. In the program in Figure~\ref{fig:simple-fp}, the constants are:
\begin{itemize}
 \item \texttt{x} is \texttt{0.1000000000000000055511151231257827021181583404541015625}
 \item \texttt{y} is \texttt{0.200000000000000011102230246251565404236316680908203125}
 \item \texttt{w} is \texttt{0.3000000000000000444089209850062616169452667236328125}
 \item \texttt{z} is \texttt{0.299999999999999988897769753748434595763683319091796875}
\end{itemize}

The discrepancy happens in the C program because the closest floatint-point to \texttt{0.3} is smaller than the real \texttt{0.3} but the closest floating-point to \texttt{0.2} is greater than the real \texttt{0.2}, so adding the floating-points \texttt{0.1} and \texttt{0.2} results in a floating-point slightly greater than the constant floating-point \texttt{0.3}.

To address this limitation, ESBMC was extended to support floating-point arithmetic~\cite{10.1007/978-3-319-70848-5_7} but was only able to encode it using SMT solvers that offered native support for the floating-point theory, i.e., Z3 and MathSAT. A floating-point is represented in ESBMC following the IEEE-754 standard for the size of the exponent and significand precision. For example, a half-precision floating-point (16 bits) has 1 bit for the sign, 5 bits for the exponent, and 11 bits for the significand (1 hidden bit)~\cite{4610935}.

The work described in this paper, namely a new floating-point API in our SMT backend, is the natural evolution of our research: the support of floating-point arithmetic for the remaining SMT solvers in ESBMC (Boolector~\cite{NiemetzPreinerBiere-JSAT15}, Yices~\cite{DBLP:conf/cav/2014}, and CVC4~\cite{Barrett:2011:CVC:2032305.2032319}). The new floating-point API works by converting all floating-point types and operations to bit-vectors (a process called bit-blasting), thus extending the support for floating-point arithmetic to any solver that supports bit-vector arithmetic~\cite{GadelhaMMCN20}.

\subsection{Bit-blasting Floating-Point Arithmetic}

The SMT \texttt{FP} logic is an addition to the SMT standard, first proposed in 2010 by R{\"{u}}mmer and Wahl~\cite{smtFPA2010}. The current version of the theory largely follows the IEEE standard 754-2008~\cite{4610935}. It formalizes floating-point arithmetic, positive and negative infinities and zeroes, NaNs, relational and arithmetic operators, and five rounding modes: round nearest with ties choosing the even value, round nearest with ties choosing away from zero, round towards positive infinity, round towards negative infinity and round towards zero.

There exist some functionalities from the IEEE standard that are not yet supported by the \texttt{FP} logic as described by Brain et al.~\cite{7203811}; however, when encoding C programs using the \texttt{FP} logic, most of the process is a one-to-one conversion, as we described in our previous work~\cite{10.1007/978-3-319-70848-5_7}.

Encoding programs using the SMT floating-point theory has several advantages over a fixed-point encoding. However, the main one is the correct modeling of ANSI-C/C++ programs that use the IEEE floating-point arithmetic. ESBMC ships with models for most of the current C11 standard functions~\cite{C11}; floating-point exception handling, however, is not yet supported.

The encoding algorithms, however, can be very complex, and it is not uncommon to see the SMT solvers struggling to support every corner case~\cite{z3-fp-bug,cvc-fp-bug}. Currently, various SMT solvers support the SMT floating-point theory, e.g., Z3~\cite{Z08}, MathSAT~\cite{Cimatti:2013:MSS:2450387.2450400}, CVC4~\cite{Barrett:2011:CVC:2032305.2032319}, Colibri~\cite{colibri2020}, Solonar~\cite{solonar2020}, and UppSAT~\cite{ZeljicBWR18}; ESBMC implements the floating-point encoding for all of them using their native API. Regarding the support from the solvers, Z3 implements all operators, MathSAT implements all but two: \texttt{fp.rem} (remainder operator) and \texttt{fp.fma} (fused multiply-add) and CVC4 implements all but the conversions to other sorts.

The three solvers offer two (non-standard) functions to reinterpret floating-points to and from bit-vectors: \texttt{fp\_as\_ieeebv} and \texttt{fp\_from\_ieeebv}, respectively. These functions can be used to circumvent any lack of operators, and only require the user to write the missing operators. Note that this is different from converting floating-points to bit-vectors and vice-versa: converting to bit-vectors follows the rounding modes defined by the IEEE-754 standard while reinterpreting floating-point as bit-vectors returns the bit-vector format of the floating-point. We use these functions in our backend to implement the fused multiply-add operator for MathSAT.

The implementation of the floating-point API is based on the encoding of Muller et al.~\cite{handbook-fp}, however, before we can discuss the algorithms in the floating-point API, we first need to describe the basic operations performed by most of them, the four-stage pipeline~\cite{BrainSS19}: unpack, operate, round, and pack.

\begin{enumerate}
 \item \textit{Unpack stage}: the floating-point is split into three bit-vectors, one for the sign, one for the exponent, and one for the significand. In our floating-point API, the unpack operation also adds the hidden bit to the significand, unbias the exponent. It offers an option to normalize subnormals exponents and significands if requested.

\item \textit{Operate stage}: in this stage, conversion and arithmetic operations are performed in the three bit-vectors. Depending on the operation, the bit-vectors need to be extended, e.g., during a fused multiply-add operation, the significand has length \texttt{2 * sb + 3}, and the exponent has length \texttt{eb + 2}.

\item \textit{Round stage}: since the previous stage was performed using extended bit-vectors, this stage needs to round the bit-vectors back to the nearest representable floating-point of the target format. Here, \textit{guard} and \textit{sticky} bits in the significand are used to determine how far the bit-vector is from the nearest representable, and the rounding mode is used to determine in which direction the floating-point will be rounded. The exponent bit-vector is also checked for under- or overflow when rounding, to create the correct floating-point, e.g., infinity might be created if the exponent is too large for the target format.

\item \textit{Pack stage}: in the final stage, the three bit-vectors are concatenated to form the final floating-point.
\end{enumerate}

The four-stage pipeline will be used when performing operations with the floating-points. We grouped the operations into seven groups: sorts constructors, rounding modes constructors, value constructors, classification operators, comparison operators, conversion operators, and arithmetic operators.

In the three constructors groups (sorts, rounding modes, and value), the floating-points are encoded using bit-vectors:

\noindent\textbf{Sorts constructors.} The sorts follow the definitions in Table~\ref{table:back-fp} for the bit-vector sizes. We do not support the 80-bit long double extended precision format used in some processors~\cite{Goldberg91whatevery}; instead, we use 128 bits for quadruple precision.

\noindent\textbf{Rounding mode constructors.} The floating-point API supports all rounding modes described in Section~\ref{back:fp}, even though the C standard does not support RNA~\cite{C11}. These are encoded as 3-bits long bit-vectors.

\noindent\textbf{Value constructors.} Floating-point literals, plus and minus infinity, plus and minus zeroes and NaNs can be created. For the later, the same NaN is always created (positive, the significand is $000\ldots01$). All values are bit-vectors with total length \texttt{1 + eb + sb}, where \texttt{eb} is the number of exponent bits and \texttt{sb} is the number of significand bits. All algorithms in the floating-point API assume one hidden-bit in the significand.

The remaining four operators groups use at least one of the stages in the pipeline to reason about floating-points:

\noindent\textbf{Classification operators.} Algorithms to classify normals, subnormals, zeros (regardless of sign), infinities (regardless of sign), NaNs, and negatives and positives. The operators work by unpacking the floating-point and comparing the bit-vectors against the definitions.

\noindent\textbf{Comparison operators.} The operators ``greater than or equal to'', ``greater than'', ``less than or equal to'', ``less than'', and ``equality'' are supported. The first three are written in terms of the last two. All of them evaluate to false if one of their arguments is NaN; this check is done using the NaN classification operator.

\noindent\textbf{Conversion operators.} The floating-point API can convert:
\begin{itemize}
 \item Floating-points to signed bit-vectors and floating-points to unsigned bit-vectors: converts the floating-point to bit-vectors always rounding towards zero. These operations generate a free variable if it can not represent the floating-point using the target bit-vector, i.e., if the floating-point is out-of-range, $\pm$NaN or $\pm$ infinity. Minus zero is converted to zero.

\item Floating-points to another floating-point: converts the floating-point to a different format using a rounding mode. $\pm$NaN, $\pm$infinity, and $\pm$zeroes are always convertible between floating-points, but converting between formats might create $\pm$infinity if the target format can not represent the original floating-point.

\item Signed bit-vectors to floating-points and unsigned bit-vectors to floating-points: converts bit-vectors to the nearest representable floating-point, using a rounding mode. It might create $\pm$infinity if the target format can not represent the original bit-vector.
\end{itemize}

\noindent\textbf{Arithmetic operators.} The floating-point API implements:
\begin{itemize}
 \item \textit{Absolute value operator}: sets the sign bit of the floating-point to
\texttt{0}.

 \item \textit{Negation operator}: flips the sign bit of the floating-point.

 \item \textit{Addition operator}: the significands are extended by 3 bits to perform the addition and the exponent are extended by 2 bits to check for overflows. The algorithm first aligns the significands then it adds them.

 \item \textit{Subtraction operator}: negates the right-hand side of the expression and
uses the addition operator, i.e., $x - y = x + (-y)$.

 \item \textit{Multiplication operator}: the length of the significand bit-vectors are
doubled before multiplying them, and the exponents are added
together. The final sign bit is the result of xor'ing the sign of both operands
of the multiplication.

 \item \textit{Division operator}: the length of both significand and exponent are
extended by 2 bits, then bit-vector subtractions are used to calculate the
target significand and exponent.

 \item \textit{Fused multiply-add}: the significand is extended to length
\texttt{2 * sb + 3} to accommodate both the multiplication and the addition, and
the exponent is extended by 2 bits. The first two operands are multiplied, and
the result is aligned with the third operand before adding them.

 \item \textit{Square root operator}: neither the significand nor the exponent is
extended since the result always fits the original format and can never
underflow or overflows as per the operator definition. Here,
$\sqrt{x} = l * 2^d$, where the final exponent $d$ is half the unbiased
exponent minus the leading zeros, and $l$ is calculated using a restorative
algorithm~\cite[Chapter 10]{handbook-fp}.
\end{itemize}

All operators but the absolute value and negation handle special values
($\pm$NaN, $\pm$infinity, and $\pm$zeroes) before performing the operations,
e.g., in the multiplication operator, if the left-hand side argument is positive
infinity, the result is NaN if the right-hand side argument is \texttt{0}; otherwise, the result is an infinity with the right-hand side argument sign.
Furthermore, all arithmetic operations in the floating-point API that take
more than one floating-point as an argument assume that the floating-points have
the same format. This assumption is not a problem when converting C programs, as type promotion
rules already ensure this pre-condition~\cite{C11}.

A detailed table with all the supported features of the floating-point API, and
the comparison with the features from other solvers can be found in
Appendix~\ref{appendix:fp-support}.

\section{Experimental Evaluation}

Our experimental evaluation consists of three parts. In Section~\ref{incr-benchmarks-description}, we present the benchmarks used to evaluate the implementation of our floating-point API. In Section~\ref{incr-fp-api}, we compare the verification results of the new floating-point API in ESBMC using several solvers. In Section~\ref{tools}, we compare the best solver found in Section~\ref{incr-fp-api} against all the tools that competed in the \textit{ReachSafety-Floats} sub-category in SV-Comp 2020. Our experimental evaluation aims to answer two research questions:
\begin{tcolorbox}
\begin{enumerate}
\item[\textbf{RQ1}] \textbf{(Soundness and completeness)} Is our floating-point API sound and complete?
\item[\textbf{RQ2}] \textbf{(Performance)} How does the implementation of our floating-point API compare to other software verifiers?
\end{enumerate}
\end{tcolorbox}

\subsection{Experimental Setup} \label{incr-benchmarks-description}

We evaluate our approach using all the verification tasks in SV-COMP 2020~\cite{svcomp2020}. In particular, we considered $466$ benchmarks for the sub-category \textit{ReachSafety-Floats}, described as \textit{``containing tasks for checking programs with floating-point arithmetics''}.

The \textit{ReachSafety-Floats} sub-category is part of the \textit{ReachSafety} category. In this category, a function call is checked for reachability; the property is formally defined in the competition as \texttt{G ! call(\_\_VERIFIER\_error())} or \textit{``The function \texttt{\_\_VERIFIER\_error()} is not called in any finite execution of the program''}.

We have implemented our floating-point API in ESBMC.
We run ESBMC on each benchmark in that sub-category once per solver, with the
following set of options:
\texttt{--no-div-by-zero-check}, which disables the division by zero check (an SV-COMP requirement);
\texttt{--incremental-bmc}, which enables the incremental BMC;
\texttt{--unlimited-k-steps}, which removes the upper limit of iteration steps in the incremental BMC algorithm;
\texttt{--floatbv}, which enables SMT floating-point encoding;
\texttt{--32}, which assumes a 32 bits architecture;
and
\texttt{--force-malloc-success}, which forces all dynamic allocations succeed to (also an SV-COMP requirement).
We also disable pointer safety checks and array bounds check (\texttt{--no-pointer-check},
\texttt{--no-bounds-check}) as, per the competition definition, these benchmarks
only have reachability bugs. Finaly, in order to select an SMT
solver for verification, the options \texttt{--boolector}, \texttt{--z3},
\texttt{--cvc}, \texttt{--mathsat}, and \texttt{--yices} are used.

All experiments were conducted on our mini cluster at the University of
Manchester, UK. The compute node used are equipped with
Intel(R) Xeon(R) CPU E5-2620 v4 @ 2.10GHz and $180$GB of RAM, where nine
instances of ESBMC were executed in parallel.
For each benchmark, we set time and memory limits of $900$ seconds and $15$GB,
respectively, as per the competition definitions. We, however, do not present
the results as scores (as it is done in SV-COMP) but show the number of correct
and incorrect results, and the verification time.

\subsection{Floating-Point API evaluation} \label{incr-fp-api}

\begin{figure}[!ht]
\includegraphics[width=1\textwidth]{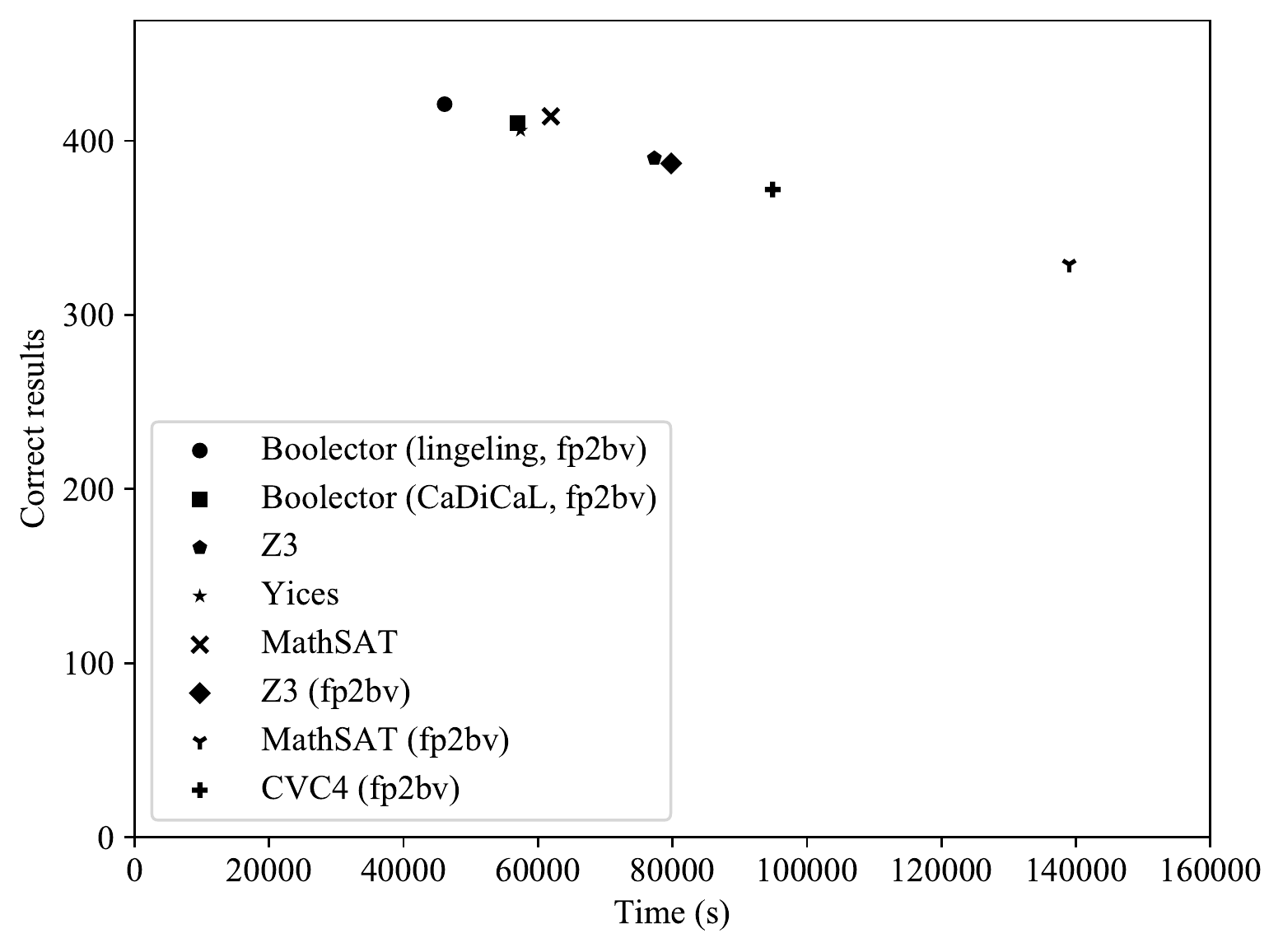}
\caption{\textit{ReachSafety-Floats} results for each solver, using the incremental BMC.
The ``fp2bv'' next to the solver name means that our floating-point API was
used to bit-blast floating-point arithmetic.}
\label{figure:fps}
\end{figure}

Figure~\ref{figure:fps} shows the number of correctly verified programs out of
the $466$ benchmarks from the \textit{ReachSafety-Floats} sub-category, using
several solvers and how long it took to complete the verification. There exists
no case where ESBMC reports an incorrect result.

Boolector (lingeling, fp2bv) reports the highest number of correct results ($421$), followed by MathSAT using their native floating-point API ($414$). This evaluation produced a slightly better result than our previous one of these solvers, where MathSAT was able to solve floating-point problems quickly but suffered slowdowns in programs with arrays~\cite{10.1007/978-3-319-70848-5_7}. MathSAT (fp2bv) presented the fewest number of correct results ($329$).

The results show that Z3 with its native floating-point API
and Z3 with our fp2bv API produce very similar results: $390$ and $387$, respectively;
this result is expected since our fp2bv API is heavily based on the bit-blasting performed by Z3 when
solving floating-points. The number of variables and clauses generated in the
CNF format, when using Z3 with its native floating-point API, is 1\%-2\% smaller
than the number generated when using our fp2bv API. The smaller number explains
the slightly better results: we assume this is the result of optimizations when
Z3 performs the bit-blasting internally.

MathSAT results show that their API can solve $85$ more benchmarks than MathSAT (fp2bv) within time and memory limits. These benchmarks contain chains of multiplications. They thus require a high computational effort during the propositional satisfiability search. Given that we replace all higher-level operators by bit-level circuit equivalents (bit-blasting), we end up destroying structural word-level information in the problem formulation. Therefore, these results lead us to believe that the MathSAT ACDL algorithm is somehow optimized for FP operations; unfortunately, MathSAT is a free but closed source tool, so we cannot confirm this.

The total verification time for each solver is also illustrated in Figure~\ref{figure:fps}, and again Boolector (lingeling, fp2bv) was the faster solver, thereby solving all programs in $46100$ seconds. It is followed by Boolector (CaDiCal, fp2bv) with $56900$, and Yices (fp2bv) with $57400$ seconds. Overall, Boolector (lingeling, fp2bv) presented the best results. It correctly verified more programs while also being the faster solver, almost $20$\% faster than the second faster solver, which is also Boolector but with a different SAT solver (CaDiCaL).

\begin{tcolorbox}
ESBMC produced no incorrect result in this evaluation, which partially answers
\textbf{RQ1}: although we can not formally prove that our algorithm is sound and complete, empirical evidence suggests it.
\end{tcolorbox}

\subsection{Comparison to other Software Verifiers}
\label{tools}

\begin{figure}[!ht]
\includegraphics[width=1\textwidth]{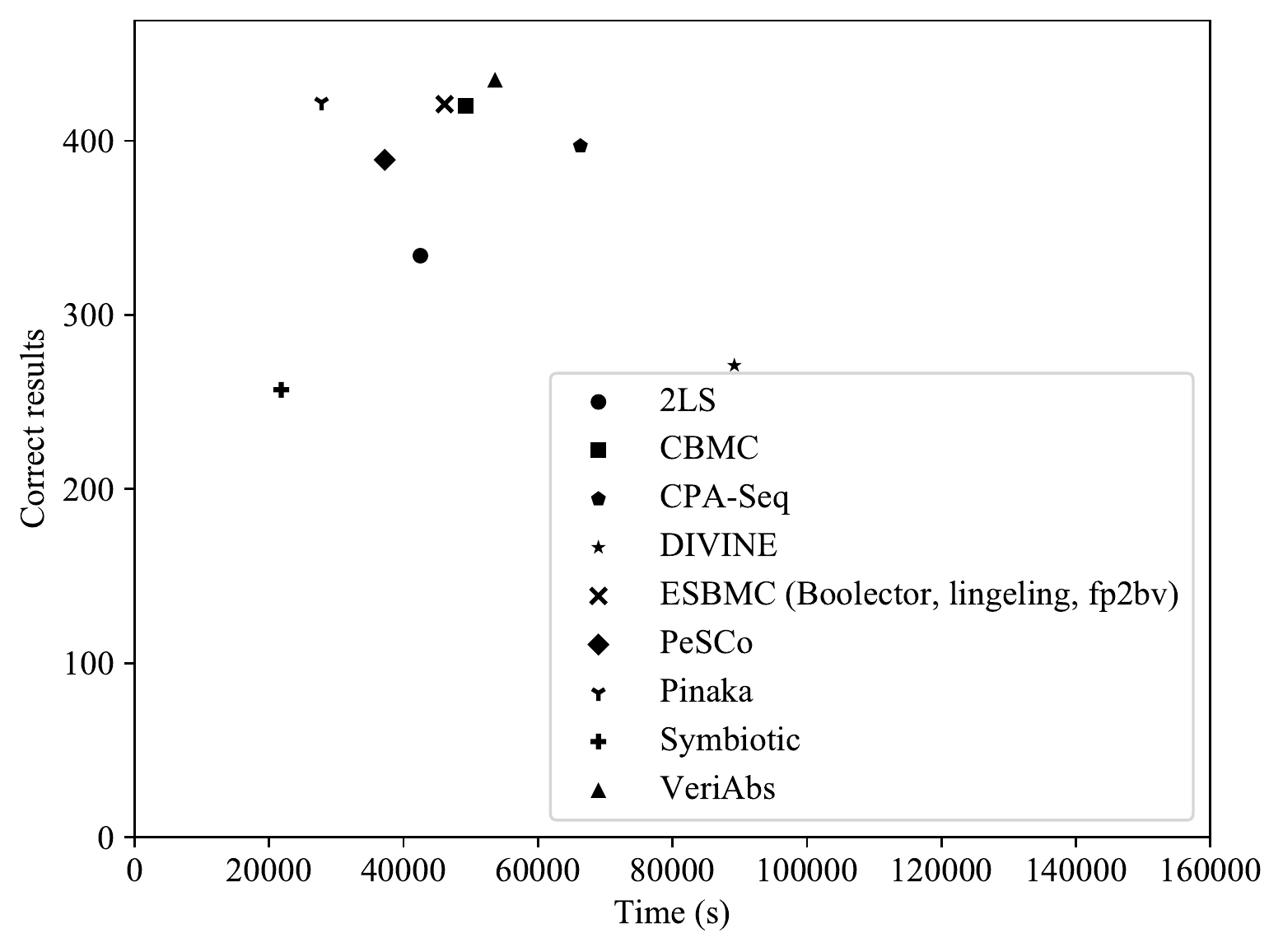}
\caption{\textit{ReachSafety-Floats} results from our best approach against tools from
SV-COMP 2020.}
\label{figure:fps-tools}
\end{figure}

We compare the implementation of our floating-point API
with other software verifiers: 2LS~\cite{10.1007/978-3-319-89963-3_24},
CBMC~\cite{CBMC:2012}, CPA-Seq~\cite{cpachecker}, DIVINE~\cite{divine4},
PeSCo~\cite{RichterW19}, Pinaka~\cite{ChaudharyJ19}, Symbiotic~\cite{ChalupaVJSS17},
VeriAbs~\cite{ChimdyalwarDCSK17}. Figure~\ref{figure:fps-tools} illustrates the
\textit{ReachSafety-Floats} results from our best approach against tools that
participated in SV-COMP 2020. In particular, we have used the binary and scripts
of these tools that are available at the SV-COMP 2020 website under ``Participating Teams''.\footnote{\url{https://sv-comp.sosy-lab.org/2020/systems.php}}
Overall, VeriAbs achieved the highest number of correct
results ($435$) in $53600$\,s followed by Pinaka ($422$) with $27800$\,s,
ESBMC ($421$) with $46100$\,s, and CBMC ($420$) with $49200$\,s.

VeriAbs can verify C programs with floating-points via abstraction using SAT solvers.
In particular, VeriAbs replaces loops in the original code by abstract loops of small known bounds;
it performs value analysis to compute loop invariants and then applies an iterative refinement using \textit{k}-induction.
The VeriAbs tool uses CBMC as its backend to prove properties and find errors, which thus
allows VeriAbs to verify C programs with floating-points. By contrast, ESBMC
uses an iterative technique and verifies the program for each unwind bound until it exhausts the time or memory limits.
Intuitively, ESBMC can either find a counterexample with up to $k$ loop unwinding
or fully unwinds all loops using the same unwinding bound so that it can provide a correct result.
ESBMC also relies on SMT solvers to check the satisfiability of the verifications conditions that
contain floating-point arithmetic.

Pinaka verifies C programs using CBMC, but it relies on an incremental SAT solving coupled with eager state infeasibility checks. Additionally, Pinaka extends CBMC to support both Breadth-First Search and Depth-First Search as state exploration strategies along with partial and full incremental modes. Here we have not evaluated the SMT incremental mode implemented in ESBMC since this feature is currently supported for the SMT solver Z3 only. Other SMT solvers do support incremental solving, but ESBMC does not provide support for incremental solving for other SMT solvers yet.

CBMC~\cite{Clarke04} implements a bit-precise decision procedure for the theory of floating-point arithmetic~\cite{Brain2014}.
Both VeriAbs and Pinaka rely on CBMC to verify the underlying C program using that decision procedure.
ESBMC originated as a fork of CBMC in 2008 with an improved SMT backend~\cite{CordeiroFM09} and
support for the verification of concurrent programs using an explicit interleaving approach~\cite{CordeiroF11}.
CBMC uses SAT solvers as their primary engine but offers support for the generation of an SMT formula for an external SMT solver. ESBMC
supports SMT solvers directly, through their APIs, along with the option to output SMT formulae.
As a result, the main difference between CBMC and ESBMC here relies on the encoding and checking of the
verification conditions that contain floating-point arithmetic.

\begin{tcolorbox}
These results answer our \textbf{RQ2}: our floating-point API is on par with other
state-of-the-art tools. VeriAbs and Pinaka implement several heuristics to
simplify the check for satisfiability using CBMC, while ESBMC used an
incremental approach produced close results. ESBMC was also slightly faster
and provided a few more results than CBMC, which lead us to believe that our
tool would also greatly benefit VeriAbs and Pinaka if used as backend.
\end{tcolorbox}

\section{Related Work}

Several symbolic execution tools try to verify programs with floating-point arithmetic by employing different strategies. CoverMe~\cite{Fu:2017:AHC:3062341.3062383} reformulates floating-point constraints as mathematical optimization problems and uses a specially built solver called XSat~\cite{Fu2016XSatAF} to check for satisfiability. Pex~\cite{Tillmann:2008:PWB:1792786.1792798} uses a similar approach and reasons for floating-point constraints as a search problem, and they are solved by using meta-heuristics search methods. FPSE~\cite{Botella:2006:SEF:1133626.1133628} models floating-point arithmetic by using an interval solver over real arithmetic combined with projection functions. 

HSE~\cite{Quan:2016:HSE:2950290.2983966} extends KLEE~\cite{Cadar:2008:KUA:1855741.1855756} to execute the program and convert floating-points into bit-vectors symbolically. It then uses SMT solvers to reason about satisfiability. Astr\'ee  is a static analysis tool that considers all possible rounding errors when verifying C programs with floating-point numbers~\cite{abs-cs-0701193}. It has been applied to verify embedded software in the flight control software of the Airbus.

Bounded model checkers have also been applied to verify programs with
floating-point arithmetic: CBMC~\cite{Clarke04} and
2LS~\cite{DBLP:journals/corr/SchrammelKBMTB14} convert floating-point operations
to bit-vectors and use SAT solvers to reason about satisfiability.
CPBPV~\cite{Collavizza:2014:GTC:2593735.2593737} uses bounded model checking
combined with their FPCS~\cite{Michel:2001:SCO:647488.726803} interval solver to
generate tests that violate output constraints in the program.

Brain et al.~\cite{BrainSS19} describe an approach called SymFPU for handling the theory of floating-point by reducing it to the theory of bit-vectors. In particular, the authors describe a library of encodings, which can be included in SMT solvers to add support for the theory of floating-point by taking into account floating-point reasoning and the fundamentals of circuit design. Brain et al. have integrated SymFPU into the SMT solver CVC4 and evaluate it using a broad set of benchmarks; they conclude that SymFPU+CVC4 can substantially out-performs all previous systems despite using a straightforward bit-blasting approach for floating-point problems. We could not compare our approach against SymFPU because of bugs in the CVC4 C API; we contacted the author, and we will create bug reports about the issues we identified.

\section{Conclusions}

We have described our new SMT floating-point API, which bit-blasts floating-point arithmetic and extends the floating-point support for SMT solvers that only support bit-vector arithmetic. The floating-point API was implemented in the SMT backend of ESBMC. Our experimental results show that Boolector (with lingeling as SAT solver)  presented the best results: the highest number of correct results within the shortest verification time. We also show that our floating-point API implemented in ESBMC is on par with other state-of-the-art software verifiers. VeriAbs and Pinaka implement several heuristics to simplify the check for satisfiability using CBMC, while ESBMC with a straightforward incremental approach produced close results.

ESBMC was already extensively used to verify digital systems~\cite{AbreuGCFS16,BessaICF16,BessaIPCF17}. However, these projects were limited to fixed-point arithmetic; supporting floating-point encoding will allow researchers to expand their activities in the scientific community. The extensive evaluation performed during the development of these technologies also identified areas to be improved in the solvers and other verification tools. In particular, we submitted patches to Z3 to optimize the generation of unsigned less-than operations during the bit-blast of floating-points\footnote{\url{https://github.com/Z3Prover/z3/pull/1501}} (accepted, part of Z3 4.6.1). We also reported bugs to both CBMC\footnote{\url{https://github.com/diffblue/cbmc/issues/1944}} and MathSAT, concerning floating-point arithmetic issues, which were later confirmed by the developers.



\clearpage

\appendix

\section{Support for the \texttt{FP} logic}
\label{appendix:fp-support}

\begin{longtable}{|l||c|c|c||c|}
      \hline
                                        & Z3      & MathSAT  & CVC4            & ESBMC   \\
      SMT \texttt{FP} operations        & v4.7.1  & v5.5.1   & v1.6-prerelease & FP API  \\ \hline
      Create floating point sort        & $\surd$ & $\surd$  & $\surd$         & $\surd$ \\ \hline
      Create rounding mode sort         & $\surd$ & $\surd$  & $\surd$         & $\surd$ \\ \hline
      Create floating point literal     & $\surd$ & $\surd$  & $\surd$         & $\surd$ \\ \hline
      Create plus and minus infinity    & $\surd$ & $\surd$  & $\surd$         & $\surd$ \\ \hline
      Create plus and minus zeroes      & $\surd$ & $\surd$  & $\surd$         & $\surd$ \\ \hline
      Crete NaN                         & $\surd$ & $\surd$  & $\surd$         & $\surd$ \\ \hline
      Absolute value operator           & $\surd$ & $\surd$  & $\surd$         & $\surd$ \\ \hline
      Negation operator                 & $\surd$ & $\surd$  & $\surd$         & $\surd$ \\ \hline
      Addition operator                 & $\surd$ & $\surd$  & $\surd$         & $\surd$ \\ \hline
      Subtraction operator              & $\surd$ & $\surd$  & $\surd$         & $\surd$ \\ \hline
      Multiplication operator           & $\surd$ & $\surd$  & $\surd$         & $\surd$ \\ \hline
      Division operator                 & $\surd$ & $\surd$  & $\surd$         & $\surd$ \\ \hline
      Fused multiply-add operator       & $\surd$ & $\times$\footnote{In ESBMC, the fused multiply-add operation uses the bit-blasting API when using MathSAT.} & $\surd$ & $\surd$ \\ \hline
      Square root operator              & $\surd$ & $\surd$  & $\surd$         & $\surd$ \\ \hline
      Remainder operator                & $\surd$ & $\times$ & $\surd$         & $\times$ \\ \hline
      Rounding to Integral operator     & $\surd$ & $\surd$  & $\surd$         & $\surd$ \\ \hline
      Minimum operator                  & $\surd$ & $\surd$  & $\surd$         & $\times$ \\ \hline
      Maximum operator                  & $\surd$ & $\surd$  & $\surd$         & $\times$ \\ \hline
      Less than or equal to operator    & $\surd$ & $\surd$  & $\surd$         & $\surd$ \\ \hline
      Less than operator                & $\surd$ & $\surd$  & $\surd$         & $\surd$ \\ \hline
      Greater than or equal to operator & $\surd$ & $\surd$  & $\surd$         & $\surd$ \\ \hline
      Greater than operator             & $\surd$ & $\surd$  & $\surd$         & $\surd$ \\ \hline
      Equality operator                 & $\surd$ & $\surd$  & $\surd$         & $\surd$ \\ \hline
      IsNormal check                    & $\surd$ & $\surd$  & $\surd$         & $\surd$ \\ \hline
      IsSubnormal check                 & $\surd$ & $\surd$  & $\surd$         & $\surd$ \\ \hline
      IsZero check                      & $\surd$ & $\surd$  & $\surd$         & $\surd$ \\ \hline
      IsInfinite check                  & $\surd$ & $\surd$  & $\surd$         & $\surd$ \\ \hline
      IsNaN check                       & $\surd$ & $\surd$  & $\surd$         & $\surd$ \\ \hline
      IsNegative check                  & $\surd$ & $\surd$  & $\surd$         & $\surd$ \\ \hline
      IsPositive check                  & $\surd$ & $\surd$  & $\surd$         & $\surd$ \\ \hline
      Convert to FP from real           & $\surd$ & $\surd$  & $\times$        & $\times$ \\ \hline
      Convert to FP from signed BV      & $\surd$ & $\surd$  & $\times$        & $\surd$ \\ \hline
      Convert to FP from unsigned BV    & $\surd$ & $\surd$  & $\times$        & $\surd$ \\ \hline
      Convert to FP from another FP     & $\surd$ & $\surd$  & $\times$        & $\surd$ \\ \hline
      Convert to unsigned BV from FP    & $\surd$ & $\surd$  & $\times$        & $\surd$ \\ \hline
      Convert to signed BV from FP      & $\surd$ & $\surd$  & $\times$        & $\surd$ \\ \hline
      Convert to real from FP           & $\surd$ & $\surd$  & $\times$        & $\times$ \\ \hline
                                        & Z3      & MathSAT  & CVC4            & ESBMC   \\
      SMT \texttt{FP} operations        & v4.7.1  & v5.5.1   & v1.6-prerelease & FP API  \\ \hline
      Convert to IEEE BV from FP\footnote{Not part of the SMT \texttt{FP} logic.\label{appendix:ref-ieee}}         & $\surd$ & $\surd$  & $\surd$         & $\surd$ \\ \hline
      Convert to floating-point from IEEE BV\textsuperscript{\ref{appendix:ref-ieee}}       & $\surd$ & $\surd$  & $\surd$         & $\surd$ \\ \hline
  \caption{Support in each SMT solver and in the ESBMC floating-point API for the operations
described in the SMT \texttt{FP} logic. A $\surd$ indicates a supported feature
while $\times$ indicates an unsupported feature.}
  \label{table:fp-support}
\end{longtable}

\end{document}